\documentstyle[12pt]{article}

\def\R{{\rm I\kern-.20em R}}
\def\N{{\rm I\kern-.20em N}}

\newcommand{\beq}{\begin{equation}}
\newcommand{\eeq}{\end{equation}}

\begin{document}
\baselineskip=20pt
\pagestyle{plain}

\title{On Foundation of the Generalized Nambu \\ Mechanics \\
(Second Version)}
\author{
Leon Takhtajan \\
Department of Mathematics \\
State University of New York at Stony Brook \\
Stony Brook, NY 11794-3651 \\
}
\maketitle

\begin{center}{\large\bf Abstract} \end{center}

\begin{quote}

We outline basic principles of canonical formalism for
the Nambu mechanics---a generalization of Hamiltonian
mechanics proposed by Yoichiro Nambu in 1973.
It is based on the notion of Nambu bracket, which
generalizes the Poisson bracket---a ``binary'' operation on
classical observables on the phase space, to the ``multiple''
operation of higher order $n \geq 3$. Nambu dynamics is described by the phase
flow given by Nambu-Hamilton equations of motion---a system of ODE's
 which involves $n-1$ ``Hamiltonians''. We introduce the fundamental identity
for the Nambu bracket---a generalization of the Jacobi
identity, as a consistency condition for the dynamics. We show that Nambu
bracket structure defines an hierarchy of infinite families of ``subordinated''
structures of lower order, including Poisson bracket structure, which satisfy
certain matching conditions.
The notion of Nambu bracket enables to define Nambu-Poisson manifolds---phase
spaces for the Nambu mechanics, which turn out to be more `rigid'' than
Poisson manifolds---phase spaces for the Hamiltonian mechanics.
We introduce the analog of the action form and the
action principle for the Nambu mechanics. In its formulation
dynamics of loops ($n-2$-dimensional chains for
the general $n$-ary case) naturally appears.
We discuss several approaches to the quantization of Nambu
mechanics, based on the deformation theory, path integral
formulation and on Nambu-Heisenberg ``commutation'' relations. In
the latter formalism we present explicit representation of Nambu-Heisenberg
relation in the $n=3$ case. We emphasize the role ternary and higher order
algebraic operations, and mathematical structures related to them,
play in passing from Hamilton's to Nambu's dynamical
picture.
\end{quote}

\section{Introduction}

In 1973 Nambu proposed a profound generalization of classical
Hamiltonian mechanics \cite{Nambu}. In his formulation
a triple (or, more generally, $n$-tuple) of ``canonical'' variables
replaces a canonically conjugated pair in the Hamiltonian formalism
and ternary (or, more generally, $n$-ary) operation---the Nambu
bracket---replaces the usual Poisson bracket. Dynamics, according to Nambu,
is determined by Nambu-Hamilton equations of motion, which use two
(or, more generally, $n-1$) ``Hamiltonians'' and replace canonical Hamilton
equations. Corresponding phase flow preserves the phase volume so that the
analog of the Liouville theorem is still valid, which is fundamental
for the formulation of statistical mechanics \cite{Nambu}.

Nambu's proposal was partially analyzed in papers \cite{BF,MS}.
In \cite{BF} it was shown that particular example of Nambu mechanics
can be treated as a six-dimensional degenerate Hamiltonian system with
three constraints and a Lagrangian being linear in velocities. In \cite{MS}
it was shown that one can use a four-dimensional phase space as well.
However, until recently, there were no attempts to formulate the basic
principles of Nambu mechanics in the invariant geometrical form
similar to that of Hamiltonian mechanics \cite{Ar}. In this paper we develop
the basics of such formalism and display novel mathematical structures which
might have physical significance.

We start by formulating the fundamental identity (FI)
for the Nambu bracket as a consistency condition for the
Nambu's dynamics. As a corollary, it yields the analog of the Poisson
theorem that Poisson bracket of integrals of motion is again an integral
of motion. Based on FI, we introduce the notion of Nambu-Poisson manifolds,
which play the same role in Nambu mechanics that
Poisson manifolds play in Hamiltonian mechanics.

We show that Nambu bracket structure contains an infinite family of
``subordinated'' Nambu structures of lower degree, including Poisson
bracket structure, with certain matching conditions. This implies that
Nambu bracket structure is, in a certain sense, more ``rigid'' then the
Poisson bracket structure. It can be seen explicitly by
showing that FI imposes rather strong constraints
on possible forms of Nambu bracket. In addition to
quadratic differential equations,
it also introduces an overdetermined system of quadratic algebraic equations
for the Nambu bracket tensor.
This is the novel feature in comparison with the Poisson bracket case,
where we have differential constraints only. Additional algebraic
requirements are ``responsible'' for the above-mentioned
``rigidity'' of the Nambu bracket structure in comparison with the
Poisson bracket structure.
Specifically, whereas any
skew-symmetric constant $2$-tensor yields a Poisson bracket, this is
no longer true for the Nambu bracket. This manifests rather
special nature of Nambu mechanics. From our point of view such
exclusiveness should be considered as an advantage of the theory
rather then of the contrary.

Next, we develop canonical formalism for the Nambu mechanics. It is based on
the analog of Poincar\'{e}-Cartan integral invariant---action form,
which is a differential form of degree $n-1$ on the extended phase
space. It has properties similar to those of
the usual Poincar\'{e}-Cartan invariant and enables us to formulate
principle of least action for Nambu mechanics.
Instead of all  possible ``histories'' of the Hamiltonian system considered in
action principle for classical mechanics (all paths connecting
initial and final points in the configuration space), in Nambu mechanics
one should consider all $n-1$-chains in the extended phase space
whose ``time-slices'' are closed $n-2$-chains satisfying certain boundary
conditions. We define
classical action as an integral of generalized Poincar\'{e}-Cartan
invariant over such $n-1$-chains and prove the principle of least action:
``world-sheets'' of a given $n-2$-chains
under the Nambu-Hamilton phase flow are extremals of the action.

Finally, we briefly discuss quantization of Nambu
mechanics. This  problem, first considered by Nambu \cite{Nambu},
is still outstanding. We indicate several possible approaches towards
its solution.
In particular, we construct a special representation of
Nambu-Heisenberg ``commutation'' relations, which is similar in spirit
to the representation of canonical Heisenberg commutation relations
by creation-annihilation operators in the space of states of the
harmonic oscillator. In our realization states are parametrized
by a lattice of algebraic integers in a cyclotomic field for the cubic root
of unity, whereas states of the harmonic oscillator are parametrized by
non-negative rational integers.

Now let us explain the main ideas of the paper in more detail.
We start with the simplest phase space for Hamiltonian mechanics---
a two-dimensional Euclidean space
$\R^{2}$ with coordinates $x,y$ and canonical Poisson bracket
$$ \{ f_1, f_2 \} =
\frac{\partial f_1}{\partial x}\frac{\partial f_2}{\partial y}-
\frac{\partial f_1}{\partial y}\frac{\partial f_2}{\partial x}=
\frac{\partial(f_1,f_2)}{\partial(x,y)}.$$
This bracket satisfies Jacobi identity
$$\{ f_1, \{f_2, f_3 \} \} + \{ f_3, \{f_1, f_2 \} \} +
\{ f_2, \{ f_3, f_1 \} \} = 0,$$
and gives rise to a Hamilton's equations of motion
$$\frac{df}{dt}=\{H, f \},$$
where $f$ is a classical observable---a smooth function on the
phase space---and $H$ is a Hamiltonian.

Generalization of this example leads to a concept of Poisson manifolds---
smooth manifolds endowned with a Poisson
bracket structure satisfying skew-symmetry condition,
Leibniz rule and Jacobi identity.

Canonical Nambu bracket \cite{Nambu} is defined for a triple of
classical observables on the three-dimensional phase space
$\R^{3}$ with coordinates $x,y,z$ by the following beautiful formula
$$\{ f_1, f_2, f_3 \}=\frac{\partial(f_1,f_2,f_3)}
{\partial(x,y,z)},$$
where the right-hand side stands for the Jacobian of the mapping
$f=(f_1,f_2,f_3): \R^{3} \mapsto \R^{3}$.
This formula naturally generalizes usual Poisson bracket
from binary to ternary operation on classical observables.\footnote{
M.\ Flato \cite{Flato1} informed me that, apparently, Nambu introduced this
bracket in order to develop a ``toy model'' for quarks considered as triples.}
Generalized Nambu-Hamilton
equations of motion involve two ``Hamiltonians'' $H_{1}$ and
$H_{2}$ and have the form
$$\frac{df}{dt}=\{H_1, H_2, f \}.$$
Corresponding phase flow on the phase space is divergence-free
and preserves the
standard volume form $dx \wedge dy \wedge dz$ --
analog of the Liouville theorem for Nambu mechanics \cite{Nambu}.

In Section 2 we prove that canonical Nambu bracket satisfies
the following fundamental identity (and its generalizations for
the $n$-ary case)\footnote{This relation was also independently introduced
by F.\ Bayen and M.\ Flato \cite{Flato2}.}
$$\{ \{ f_1,f_2,f_3 \},f_4,f_5 \} + \{ f_3, \{f_1, f_2, f_4 \}, f_5 \}+
\{ f_3, f_4, \{ f_1, f_2, f_5 \}\} =\{ f_1, f_2, \{f_3, f_4, f_5 \}\}.$$
This formula might have been considered as the most natural (at least from
a ``dynamical'' point of view) generalization of the
Jacobi identity. It yields (see Theorem 3) an analog of
classical Poisson theorem that Poisson bracket of two integrals
of motion is again an integral of motion. Generalized version of
FI for the $n$-ry case (where Jacobian
of the mapping $f=(f_{1}, \ldots , f_{n}): \R^{n} \mapsto \R^{n}$
defines canonical Nambu bracket of order $n$) enables us to introduce
Nambu-Poisson manifolds as a smooth manifolds
with Nambu bracket structure---
an $n$-ary operation on classical observables satisfying
skew-symmetry condition, Leibniz rule and FI.
We show that by fixing some of the arguments in the Nambu bracket
of order $n$ one can get brackets of lower order which still satisfy
fundamental identity (i.e.\ are Nambu brackets) and, in addition, satisfy
certain matching conditions for different choices of fixed arguments.
This hierarchical structure of Nambu bracket shows that it is more ``rigid''
concept yen that of Poisson bracket. We discuss explicit conditions
FI imposes on corresponding $n$-tensor of the Nambu bracket.
We show, contrary to the Poisson case, that algebraic part of FI substantially
reduces possible Nambu structures with constant skew-symmetric $n$-tensor.
Specifically, we show that bracket on the phase space
$\R^{3n}=\oplus_{i=1}^{n} \R^3$,
defined in \cite{Nambu} (and used in \cite{H}) as a direct sum of
canonical Nambu brackets on $\R^3$, does not satisfy
FI and, therefore, is not a Nambu bracket. Next, we discuss linear
Nambu brackets. We show that
they naturally lead to a new notion of Nambu-Lie ``gebras'', which
generalizes Lie algebras for the $n$-ary case.
We close Section 2 by presenting several simple examples of evolution
equations which admit Nambu formulation.

In Section 3 we introduce the analog of Poincar\'{e}-Cartan integral
invariant for Nambu mechanics---a differential form of degree
$n-1$ on the extended phase space. In the simplest three-dimensional example
described above it is given by the following $2$-form on $\R^4$
$$\omega^{(2)}=x dy \wedge dz -H_1 dH_2 \wedge dt.$$
We prove in Theorem 6 that the vector field of Nambu-Hamilton phase flow
in the extended phase space is the line field of the $3$-form
$d\omega^{(2)}$, so that integral curves are its characteristics.
We define a classical action as an integral of Poincar\'{e}-Cartan action
form over
$n-1$-chains, and in Theorem 7 prove the principle of least action. It
states that $n-1$-chains---``tubes'' of integral curves of
Nambu-Hamilton phase flow ``passing through'' a given $n-2$-chains---
are the extremals  of the action. These results generalize the basic facts
lying the foundation of Hamiltonian mechanics (see, e.g., \cite{Ar}).

In Section 4 we discuss possible approaches to the quantization of
Nambu mechanics. Though we mention those based on the
deformation theory and Feynmann path integral, our main result is the explicit
construction of a special representation of Nambu-Heisenberg commutation
relation. For the ternary case this relation has the form (see \cite{Nambu})
$$[A_1,A_2,A_3]=A_1 A_2 A_3 - A_1 A_3 A_2 + A_3 A_1 A_2 - A_3 A_2 A_1 +
A_2 A_3 A_1 -A_2 A_1 A_3 = cI,$$
where $A_1, A_2, A_3$ are linear operators, $I$ is a unit operator and $c$ is
a constant.
Let ${\bf Z}[\rho]$ be a lattice of algebraic integers in quadratic
number field ${\bf Q}[\rho]$, where $\rho^3=1$.
We prove in Theorem 8 that Nambu-Heisenberg relation can be
represented by operators $A_i$ acting in the linear space ${\bf H}=
\{|\omega\rangle,~ \omega \in {\bf Z}[\rho]\}$ by the following simple formula
$$A_1 |\omega\rangle=(\omega+1+\rho)|\omega +1\rangle,~ A_2 |\omega\rangle=
(\omega+\rho)|\omega +\rho\rangle,~
A_3 |\omega\rangle=\omega |\omega + \rho^2\rangle.$$
This realization should be compared with canonical representation of
Heisenberg commutation relations in the space of states of harmonic
oscillator given by
creation-annihilation operators. Contrary to the latter case,
our representation
does not have a vacuum vector (at least in the conventional sense) and we do
not know whether the analog of Stone-von Neumann
theorem (unitary equivalence of irreducible representations) is still true.

{\bf Acknowledgments.} I appreciate stimulating conversations and discussions
with Moshe Flato, Igor Krichever, Nicolai Reshetikhin, Han Sah and Daniel
Sternheimer. I am grateful to Larry Lambe for performing symbolic
computations in connection with a system of quadratic algebraic
equations described in Section 2.
I should also mention that Theorem 8 was obtained as early as in
1977. Its publication was deliberately postponed since at that time it
did not ``fit'' into a general picture.

This work was partially supported by the Paul and Gabriella Rosenbaum
Foundation.

\section{Nambu-Poisson Manifolds}

In Hamiltonian mechanics a smooth manifold $X$ is called a
{\it Poisson manifold} and its function ring  $A=C^{\infty}(X)$---{\it
algebra of observables,} if there exists a map $ \{ ~, ~ \}:
A \otimes A \mapsto A$ with the following properties.

1. Skew-symmetry
$$\{f_1, f_2 \}= -\{ f_2, f_1 \}, $$
for all $f_1, f_2 \in A$.

2. Leibniz rule (derivation property)
$$\{ f_1f_2, f_3 \}=f_1 \{ f_2 ,f_3 \}+ f_2 \{f_1, f_3 \}.$$

3. Jacobi identity
$$\{ f_1, \{ f_2, f_3 \} \} + \{ f_3, \{ f_1, f_2 \} \} +
\{ f_2, \{ f_3, f_1 \} \}=0,$$
for all $f_1, f_2, f_3 \in A$.

Corresponding ``binary operation'' $\{ ~,~\}$ on $A$ is called {\it Poisson
bracket} and plays a fundamental role in classical mechanics.
Namely, according to Hamilton, dynamics on the {\it phase space} $X$
is determined by a distinguished function $H \in A$---a {\it Hamiltonian},
and is described by Hamilton's equations of motion
$$\frac{df}{dt}= \{ H, f \}, ~~f \in A.$$
When solution to Hamilton's equations exists for all times $t \in \R$
and all initial data
(this is so when $X$ is compact), it defines Hamilton
phase flow $x \mapsto g^t(x),~x \in X,$ and
evolution operator $U_t: A \mapsto A$,
$$U_{t}(f)(x)=f(g^{t}(x)),~x \in X,~f \in A.$$

Hamilton's dynamical picture is consistent if and only if evolution
operator $U_t$ is an isomorphism of algebra of observables $A$.
This means that $U_t$ is an algebra isomorphism, i.e.\
$U_{t}(f_1 f_2)=U_{t}(f_1)U_{t}(f_2)$ and, in addition, preserves the Poisson
structure on $A$, i.e.\ $U_{t}(\{ f_1, f_2 \})=\{ U_{t}(f_1), U_{t}(f_2) \}$.
It is easy to see (using the standard uniqueness theorem for ODE's)
that the first property is equivalent to the Leibniz rule and
the second one is equivalent to the Jacobi identity. We summarize these
well known results as the following theorem.

{\bf Theorem 1} {\it Evolution operator in Hamilton's dynamical
picture is an isomorphism of algebra of observables $A=C^{\infty}(X)$ if and
only if the phase space $X$ is a Poisson manifold.}

The basic examples of Poisson manifolds are
given by two-dimensional phase space $X=\R^{2}$ with coordinates $x,y$
and Poisson bracket
$$\{ f_1, f_2 \} =
\frac{\partial f_1}{\partial x}\frac{\partial f_2}{\partial y}-
\frac{\partial f_1}{\partial y}\frac{\partial f_2}{\partial x}=
\frac{\partial(f_1,f_2)}{\partial(x,y)},$$
and by its generalization---
$X=\R^{2N}$ with coordinates $x_{1}, \ldots ,x_{N}, y_{1}, \ldots , y_{N}$
and Poisson bracket
$$\{f_{1},f_{2}\} =
\sum_{i=1}^{N}
(\frac{\partial f_{1}}{\partial x_{i}}
\frac{\partial f_{2}}{\partial y_{i}}-
\frac{\partial f_{1}}{\partial y_{i}}
\frac{\partial f_{2}}{\partial x_{i}}).$$
Geometrically, Poisson manifold $X$ is characterized by a
{\it Poisson tensor} $\eta$ --
a section of the exterior square $\wedge^{2}TX$ of a tangent bundle
$TX$ of $X$, which defines Poisson structure by the formula
$$\{ f_1, f_2 \} = \eta(df_1, df_2).$$
Jacobi identity is equivalent to the property that $\eta$ has
a vanishing Schouten bracket with itself.
In local coordinates $(x_1, \ldots ,x_N)$ on $X$ Poisson tensor $\eta$ is
given by
$$\eta=\sum_{i,j=1}^{N}\eta_{ij}(x) \frac{\partial}{\partial x_i} \wedge
\frac{\partial}{\partial x_j},$$
and Jacobi identity identity takes the form
$$\sum_{l=1}^{N}(\eta_{il}\frac{\partial \eta_{jk}}{\partial x_l}
+ \eta_{jl}\frac{\partial \eta_{ki}}{\partial x_l}+ \eta_{kl}
\frac{\partial \eta_{ij}}{\partial x_l})=0,$$
for all $i,j,k=1, \ldots ,N$.

Dynamics according to Nambu consists in replacing Poisson
bracket by a ternary ($n$-ary)
operation on algebra of observables $A$ and requires two
($n-1$) ``Hamiltonians'' $H_1,H_2$ ($H_1, \ldots ,H_{n-1}$)
to describe the evolution. This dynamical picture is consistent if and only if
evolution operator is an isomorphism of algebra of observables. Therefore,
we propose the following definition.

{\bf Definition 1}
Manifold $X$ is called a {\it Nambu-Poisson manifold} of order $n$ if there
exists a map $\{, \ldots , \}: A^{\otimes^n} \mapsto A$---generalized
{\it Nambu bracket} of order $n$, satisfying the following properties.

1. Skew-symmetry
$$\{ f_1, \ldots ,f_n \}=(-1)^{\epsilon(\sigma)}\{ f_{\sigma(1)}, \ldots ,
f_{\sigma(n)} \},$$
for all $f_1, \ldots ,f_n \in A$ and $\sigma \in Symm(n)$,
where $Symm(n)$ is a symmetric group of $n$ elements
and $\epsilon(\sigma)$ is the parity of a permutation $\sigma$.

2. Leibniz rule
$$\{ f_1 f_2, f_3, \ldots ,f_{n+1} \}=
f_1 \{f_2, f_3, \ldots , f_{n+1} \} +
f_2 \{ f_1, f_3, \ldots, f_{n+1} \},$$
for all $f_{1}, \ldots ,f_{n+1} \in A$.

3. Fundamental identity (FI)
\begin{eqnarray*}
\{ \{ f_1, \ldots , f_{n-1}, f_n \}, f_{n+1}, \ldots, f_{2n-1} \} +
\{ f_n, \{ f_1, \ldots, f_{n-1}, f_{n+1} \}, f_{n+2}, \ldots , f_{2n-1} \} \\
 +  \ldots + \{ f_n, \ldots ,f_{2n-2}, \{ f_1, \ldots , f_{n-1}, f_{2n-1} \}\}
 =  \{ f_1, \ldots , f_{n-1}, \{ f_n, \ldots , f_{2n-1} \}\},
\end{eqnarray*}
for all $f_1, \ldots , f_{2n-1} \in A.$

{\bf Remark 1} Nambu bracket structure of order $n$ on a phase space $X$
induces infinite family of ``subordinated'' Nambu structures
of orders $n-1$ and lower, including the family of Poisson structures.
Indeed, consider $n=3$ case and for $H \in A$
define the bracket $\{~,~\}_{H}$ on $X$ as
$$\{\psi,\phi \}_H = \{H,\psi, \phi \}$$
for all $\psi, \phi \in A$. Setting in FI $f_1=f_3=H$, we see that it turns
into the Jacobi identity for the family of brackets $\{~,~\}_H$
parametrized by observables $H$. Conversely, such family of Poisson brackets
gives rise to Nambu bracket if certain matching conditions are satisfied.
Namely, for any $\phi \in A$ define
$$D^{H}_{\phi}(f)=\{\phi,f\}_H,~f \in A,$$
which is the derivation of the Poisson bracket $\{,~,~\}_H$ (Jacobi identity).
Then the family $\{~,~\}_H,~H \in A$ of Poisson brackets on $X$ gives rize to
a Nambu bracket, defined as
$$\{f_1,f_2,f_3\}=\{f_2,f_3\}_{f_1},$$
if and only if
$$D^{H_1}_{\phi}(\{\psi,\chi\}_{H_2})=\{D^{H_1}_{\phi}(\psi),\chi \}_{H_2}
+\{\psi, D^{H_1}_{\phi}(\chi)\}_{H_2} +
\{\psi, \chi\}_{D^{H_1}_{\phi}(H_2)},~\phi,\psi,\chi \in A,$$
for all $H_1,H_2 \in A$. Indeed, it is easy to see that
this equation---a derivation property
of $D^{H_1}_{\phi}$ with respect to the whole family of Poisson brackets
$\{~,~\}_{H}$---is equivalent to FI (if one identifies $f_1=H_1, f_2=\phi,
f_3=\psi,f_4=\chi$ and $f_5=H_2$). Moreover, this condition for the case
$H_1=H_2=H$ is equivalent to the Jacobi identity for the bracket $\{~,~\}_H$.
The same is true for the general case,
where for all $H_1, \ldots H_{n-k} \in A$ the assignment
$$\{f_1, \ldots ,f_k\}_{H_1 \ldots H_{n-k}}=\{H_1, \ldots, H_{n-k},f_1, \ldots
f_k\}$$
defines a hierarchy of subordinated Nambu structures of orders $k=2, \ldots ,
n-1$, parametrized by the elements in $ \wedge^{n-k} A$.
They all satisfy FI (which follows from FI for the ``basic'' structure of
order $n$) and matching conditions of the same type as above.

Dynamics on Nambu-Poisson manifold is determined by $n-1$
functions $H_1, \ldots ,H_{n-1}$ and is described
by generalized Nambu-Hamilton equations of motion
\beq \label{N-H}
\frac{df}{dt}=\{ H_1, \ldots , H_{n-1}, f \},~f \in A.
\eeq
Corresponding Nambu-Hamilton phase flow $g^t$  defines evolution operator
$U_t,~ U_{t}(f)(x)=f(g^{t}x),~x \in X,$ for all $f \in A$.

The following theorem clarifies ``dynamical'' meaning of a concept of
Nambu-Poisson manifolds.

{\bf Theorem 2} {\it Evolution operator in Nambu's dynamical picture is an
isomorphism of algebra of observables $A=C^{\infty}(X)$ if and only if the
phase space $X$ is a Nambu-Poisson manifold.}

{\it Proof.} We need to prove that
\beq \label{Ev}
U_{t}(\{ f_1, \ldots , f_n \})=\{ U_{t}(f_1), \ldots , U_{t}(f_n) \}.
\eeq
Since (\ref{Ev}) is obviously valid at time $t=0$, it sufficient to show
that it both sides satisfy the same evolution differential equation.
Denoting by $L \in Vect(X)$ the vector field of the Nambu-Hamilton flow $g^t$,
i.e.
$$L(f)=\{ H_1, \ldots , H_{n-1}, f \},$$
we can express $t$-derivative of (\ref{Ev}) as
$$L(\{ f_1, \ldots , f_n \})=\{ L(f_1), f_2, \ldots , f_n \} +
\{ f_1, L(f_2), \ldots, f_n \} + \{ f_1, f_2, \ldots, L(f_n) \},$$
which is nothing but FI
specialized for the functions $H_1, \ldots, H_{n-1}, f_1, \ldots, f_n.$
$\Box$

{\bf Remark 2} Equivalent formulation of FI used in the previous proof
can be stated explicitly that for any elements $H_1, \ldots, H_{n-1}$ the
mapping $L: A \mapsto A$ is a derivation of the Nambu bracket.

{\bf Definition 2}
Observable $F \in A$ is called {\it
integral of motion} for Nambu-Hamilton system with Hamiltonians $H_1,
 \ldots ,H_{n-1}$, if
$$ \{ H_1, \ldots , H_{n-1}, F \}=0.$$

As an obvious corollary of FI we get the following result.

{\bf Theorem 3} {\it Nambu bracket of $n$ integrals of motion is an
integral of motion.}

{\bf Remark 3} As we have seen in Remark 1, Nambu bracket structure
of order $n$ contains an infinite family of subordinated structures of lower
degree, including Poisson bracket structure, with matching conditions between
them. Therefore one might suspect that Nambu structure should be more ``rigid''
then its Poisson counterpart. This can be seen by comparing FI of order $n$
with Jacobi identity---its special case when $n=2$. As we know, the left
hand side of Jacobi identity, considered as a map from
$A \otimes A \otimes A$ into $A$, is a deriviation with respect to every
argument. However, for the general case $n \geq 3$ the difference
between the left and right hand sides of FI, considered as a map from
$2n-1$-fold tensor product $A \otimes \ldots \otimes A$ into $A$,
is a derivation only with respect to the arguments
$f_n, \ldots ,f_{2n-1}$ and not to $f_1, \ldots, f_{n-1}$.
This is because these two groups appear in FI in a different way. Namely,
analyzing FI, it is easy to see that
observables from the first group appear only twice under the
``double Nambu bracket'', whereas there $n$ such terms for every member
in the second group\footnote{ This is not surprising since, according to
interpretation in Remark 2, members from the second
group may be considered as a Hamiltonians for Nambu-Hamilton equations of
motion, whereas members from the first group are ``just'' observables.}.
Consider, for instance, such terms for the argument $f_{2n-1}$.
Denoting by $L_{i_1 \ldots i_{n-1}}$ vector field corresponding to
Nambu-Hamilton phase flow with Hamiltonians $f_{i_1}, \ldots , f_{i_{n-1}}$,
we can arrange these
terms as a commutator $[L_{n \ldots 2n-2},L_{1 \ldots n-1}](f_{2n-1})$,
so they are still given by a vector field action. This proves the derivation
property with respect to the argument $f_{2n-1}$ (cf.\ standard
arguments in Hamiltonian mechanics \cite{Ar}). Same arguments apply to all
members of the second group, but not to those of the first group. As we shall
see below, this feature of FI implies strong constraints on the possible forms
of Nambu bracket.

Geometrically, Nambu structure of order $n$ can be realized as
$$\{ f_1, \ldots , f_n \} = \eta(df_1, \ldots, df_n),$$
where $\eta$ is a section of the $n$-fold exterior power $\wedge^{n}TX$ of
a tangent bundle $TX$. In local coordinates
$(x_1, \ldots , x_N)$ on $X$ {\it Nambu tensor} $\eta$ is given by
$$\eta=\sum_{i_{1},...,i_{n}=1}^{N} \eta_{i_1...i_n}(x) \frac{\partial}{
\partial x_{i_1}} \wedge \ldots \wedge \frac{\partial}{\partial x_{i_n}}$$
and should satisfy FI. As we mentioned earlier, FI implies strong
constraints on $n$-tensor $\eta$.

First, taking into the account Remark 3, we see that all terms containing
second derivatives of $f_1, \ldots , f_{n-1}$ should vanish.
This results in the following system of quadratic algebraic
equations
\beq \label{A1}
N_{ij} + P(N)_{ij} =0,
\eeq
for all multi-indices $ i=\{i_1, \ldots , i_n \}$ and
$j=\{j_1, \ldots ,j_n \}$ from the set $\{1, \ldots , N \}$,
where
\beq \label{A2}
N_{ij}=\eta_{i_1 \ldots i_n} \eta_{j_1 \ldots
j_n} +
\eta_{j_n i_1 i_3 \ldots i_n}\eta_{j_1 \ldots j_{n-1}i_2}+ \ldots
+ \eta_{j_n i_2 \ldots i_{n-1} i_1}\eta_{j_1 \ldots j_{n-1} i_n} -
\eta_{j_n i_2 \ldots i_n}\eta_{j_1 \ldots j_{n-1} i_1}.
\eeq
and $P$ is a permutation operator which interchanges first and $n+1$-th
indices (i.e.\ $i_1$ and $j_1$) of a $2n$-tensor $N$.

Second, all terms containing first derivatives of $f_1, \ldots , f_{2n-1}$
must vanish. This yields the
following system of quadratic differential equations
\begin{eqnarray} \label{Diff}
\sum_{l=1}^{N}(\eta_{l i_2 \ldots i_n} \frac{\partial \eta_{j_1 \ldots j_n}}
{\partial x_l} + \eta_{j_n l i_3 \ldots i_n}\frac{\partial \eta_{j_1 \ldots
j_{n-1} i_2}}{\partial x_l} + \ldots + \eta_{j_n i_2 \ldots i_{n-1}l}
\frac{\partial\eta_{j_1 \ldots j_{n-1} i_n}}{\partial x_l})=0,
\end{eqnarray}
for all indices $i_2,\ldots, i_n, j_1, \ldots , j_n=1, \ldots, N$.

Thus skew-symmetric $n$-tensor $\eta$ defines Nambu bracket
of order $n$ if and only if it satisfies equations (\ref{A1}) -- (\ref{Diff}).

This shows significant difference between
Nambu and Hamiltonian formulations---a constant skew-symmetric
$n$-tensor $\eta$ (which obviously satisfies (\ref{Diff})) for $n \geq 3$ no
longer ``automatically'' defines Nambu bracket! To do so, it must
satisfy algebraic constraints (\ref{A1}) -- (\ref{A2}).
One may wonder whether there are any solutions at all to the
algebraic-differential system (\ref{A1}) -- (\ref{Diff})?

The following geometric interpretation of tensor $N$ provides the simplest
examples.
Let $V$ be the $N$-dimensional linear space and $V^{*}$ be its dual
space. Any constant
skew-symmetric $n$-tensor $\eta$ can be interpreted as an element
in the linear space $\wedge^n V$, which we also will denote by $\eta$.
With every $\eta \in \wedge^n V$ one can associate a map
$N: \wedge^{n-1} V^{*} \mapsto \wedge^{n+1} V$, defined by the formula
$$Na=i_{a}(\eta) \wedge \eta \in \wedge^{n+1} V,$$
for all $a \in \wedge^{n-1} V^{*}$. Here $i_{a}(\eta) \in V$ is
given by $(i_{a}(\eta), v^{*})= (\eta, a \wedge v^{*})$ for any $v^{*}
\in V^{*}$ and $(~,~)$ stands for the pairing between $V$ and $V^{*}$.
It is well known that equation $N=0$ is equivalent to the condition that
element $\eta \in \wedge^n V$ is decomposable,
i.e.\ there exist $v_1, \ldots v_n \in V$ such that
$\eta=v_1 \wedge \ldots \wedge v_n$,
and it is easy to verify that in coordinates the
map $N$ is represented by the $2n$-tensor $N$ given by formula (\ref{A2}).

Thus we have proved the following result.

{\bf Theorem 4} {\it Let $V$ be a linear space. Any decomposable
element in $\wedge^n V$ endows $V$ with the structure of a Nambu-Poisson
manifold of order $n$.}

In particular, consider Nambu's original example \cite{Nambu},
when $X=\R^{n}$ is
a phase space with coordinates $x_{1}, \ldots ,x_{n}$ and
``canonical'' Nambu bracket is given by
\beq \label{CanBracket}
\{ f_1, \ldots ,f_n \}=\frac{\partial(f_1, \ldots ,f_n)}
{\partial(x_1, \ldots ,x_n)},
\eeq
where the right-hand-side stands for the Jacobian of the mapping
$f=(f_1, \ldots, f_n): \R^{n} \mapsto \R^{n}$.
{}From Theorem 4 we get

{\bf Corollary} {\it Euclidean space $\R^n$ with canonical Nambu bracket
of order $n$ is a Nambu-Poisson manifold.}

{\it Proof.} Canonical
Nambu bracket of order $n$ is given by totally anti-symmetric $n$-tensor
$\eta_{i_1 \ldots i_n}= \epsilon_{i_1 \ldots i_n}$,
which corresponds to the volume element of $\R^n$ and, therefore, is
decomposable. $\Box$

{\bf Remark 4} We define {\it canonical transformations} as a diffeomorphisms
of the phase space which preserve Nambu bracket structure.
For the Nambu's example linear canonical transformations form a group
$SL(n, \R)$.

{\bf Remark 5}
In \cite{Nambu} Nambu also considered
another example---``direct sum'' of canonical
brackets. In this case  $X=\R^{3n}=\oplus_{i=1}^n \R^3$ and the bracket is
given by the following element
$$\eta = e_1 \wedge e_2 \wedge e_3 + e_4 \wedge e_5 \wedge e_6 +
\ldots + e_{3n-2} \wedge e_{3n-1} \wedge e_{3n} \in \wedge^3 \R^{3n},$$
where $e_i,~i=1, \ldots ,3n$ is a basis in $\R^{3n}$ induced by a standard
basis in $\R^3$.
(This bracket was used in \cite{H} to write equations of motion
of a particle interacting with $SU(2)$ monopole).
However, it is easy to see that such tensor $\eta$ does not satisfy system
(\ref{A1}) and, therefore, does not
define Nambu bracket! This ``explains'' Nambu's observation  \cite{Nambu}
that linear canonical
transformations for this bracket ``decouple'', i.e.\
form a direct product
of $n$ copies of $SL(3, \R)$---a fact he considered to be rather
disappointing.

To summarize this discussion, we see that there is significant
difference between Nambu and Hamiltonian mechanics with Nambu formulation
being ``more rigid''.

The problem of constructing other examples of Nambu-Poisson manifolds
is of great importance. Not trying to address it here (but hoping to
to return to it later on), we will mention a just a few.

{\bf Remark 6}
It seems that in the constant case there should be other
examples of Nambu brackets besides those given by
decomposable tensors, since equations (\ref{A1}) do not immediately
imply that $N=0$\footnote{Larry Lambe, using symbolic
computations technique, present an explicit form of a Groebner
basis of a polynomial ideal
of all relations (\ref{A1}) in the case $n=3,~N=6$. It follows from
his analysis that in this case all solutions are represented by decomposable
tensors.}.

Another class of examples is provided by a
non-constant tensors $\eta$. As we know,
linear Poisson bracket structure (Poisson structure on linear space such
that Poisson bracket of linear functions is again linear) is equivalent
to a Lie algebra structure on the dual space.
One can ask what kind of structure does linear Nambu bracket introduce?

{\bf Definition 3} A vector space $V$ is called {\it Nambu-Lie ``gebra''}
 of order $n$ if there exists a map---{\it Nambu-Lie bracket}---$[~.,\ldots ,
.~]: \wedge^n V \mapsto V$ such that
\begin{eqnarray*}
[[v_1, \ldots , v_{n-1}, v_n], v_{n+1}, \ldots, v_{2n-1}] +
[v_n, [v_1, \ldots , v_{n-1}, v_{n+1}], v_{n+2}, \ldots, v_{2n-1}] \\
+ \ldots + [v_n, \ldots, v_{2n-2}, [v_1, \ldots, v_{n-1}, v_{2n-1}]]=
[v_1, \ldots, v_{n-1}, [v_n, \ldots, v_{2n-1}]],
\end{eqnarray*}
for all $v_1, \ldots, v_{2n-1} \in V$.

{\bf Remark 7} Definition of Nambu-Lie ``gebras'' can be stated in equivalent
form as a condition that for any $v_1, \ldots , v_{n-1} \in V$
``adjoint'' map $[v_1, \ldots , v_{n-1},~]: V \mapsto V$ is a derivation
with respect to the Nambu-Lie bracket $[~, \ldots,~]$.

{\bf Theorem 5} {\it Linear Nambu structures of order $n$ are in one-to-one
correspondence with Nambu-Lie ``gebras'' of order $n$ on the dual space.}

{\it Proof.} Fundamental identity for linear functions is nothing but
FI in the definition of Nambu-Lie ``gebras''. $\Box$

In coordinates, linear Nambu structure of order $n$ is given by
$n+1$-tensor $c_{i_1 \ldots i_n}^{k}$:
$$\{x_{i_1}, \ldots, x_{i_n}\}=\sum_{k=1}^{N}c^{k}_{i_1 \ldots i_n}x_k.$$
``Structure constants'' $c^{k}_{i_1 \ldots i_n}$ satisfy overdetermined
system of quadratic algebraic equations which follow from (\ref{A1})---(\ref
{Diff}). We are planning to analyze this interesting  structure
elsewhere.

Now we present several simple examples of dynamical systems which
admit Nambu formulation.

{\bf Example 1} It goes back to Nambu \cite{Nambu} and is given by
Euler equations for the angular momentum of a rigid body in three
dimensions. These equations admit both
Hamiltonian formulation with respect to the linear Poisson
bracket on $\R^{3} \cong su(2)^{*}$, where Hamiltonian
is given by kinetic energy and Nambu formulation with respect to the
canonical ternary Nambu bracket on $\R^{3}$ and two Hamiltonians --
kinetic energy and total angular momentum.

{\bf Example 2} Lagrange system (sometimes called Nahm's system in the
theory of static $SU(2)$-monopoles, see, e.g.,  \cite{AKC,T})
on $\R^{3}$, which is given by the following equations of motion
$$\frac{dx_1}{dt} = x_2 x_3,~ \frac{dx_2}{dt} = x_1 x_3,~ \frac{dx_3}{dt}=
x_1 x_2,$$
can be written in Nambu form
$$\frac{dx_i}{dt}=\{ H_1, H_2, x_i \},~i=1,2,3,$$
where $H_1=x_{1}^{2}-x_{2}^{2},~H_2=x_{1}^{2}-x_{3}^{2}$. Integrals
$H_1$ and $H_2$ confine the phase flow to the intersection of two quadrics
in $\R^{3}$ -- a locus of an elliptic curve so that the
system can be integrated by elliptic functions. There exists another system
with quadratic nonlinearity---so-called Halphen system, which
is related to the Lagrange system (see, e.g., \cite{AKC,T}). It can
be integrated in terms of modular forms (see, e.g., \cite{T})
and does not admit global (single-valued) integrals of motion. However, this
system has two multi-valued integrals which play the role of two Hamiltonians
in Nambu formulation \cite{C}.

{\bf Example 3} Let $X=\R^n$ be the phase space with canonical Nambu bracket
of order $n$ and let the elementary symmetric functions
of $n$ variables $x_1, \ldots, x_n$ to be the Hamiltonians:
$$H_1=s_1, \ldots, H_{n-1}=s_{n-1},$$
where $(x-x_1) \ldots (x-x_n)=x^n - s_1 x^{n-1} \ldots \pm s_n.$
Nambu-Hamilton equations of motion
$$\frac{dx_i}{dt}=\{ H_1, \ldots, H_{n-1}, x_i \},~i=1, \ldots, n,$$
can be written as
$$\frac{dx_i}{dt}=\frac{\partial^n f}{\partial x_{i}^{n}}, ~~i=1, \ldots, n,$$
where
$$f(x_1, \ldots, x_n)=\prod_{1 \leq i<j \leq n}(x_i - x_j),$$
which might suggest a generalization of a gradient flows.

\section{Canonical Formalism and Action Principle}

Here we extend canonical formalism of Hamiltonian mechanics,
based on Poincar\'{e}-Cartan integral invariant and on the principle of least
action (see, e.g., \cite{Ar}), to the case of Nambu
mechanics. We illustrate essential features of our approach on the
simplest example of Nambu-Poisson manifold
$X=\R^{3}$ with canonical Nambu bracket.

Let $\tilde{X}=\R^{4}$ be the extended phase space with coordinates
$x,y,z,t$.

{\bf Definition 4}
The following $2$-form $\omega^{(2)}$ on $\tilde{X}$
\beq \label{P-C}
\omega^{(2)}=x dy \wedge dz - H_{1}dH_{2} \wedge dt,
\eeq
is called {\it generalized Poincare-Cartan integral invariant---action
form} for Nambu mechanics (cf.\ \cite{Ar}).

Consider Nambu-Hamilton equations with Hamiltonians $H_1$ and $H_2$
$$\frac{df}{dt}=\{H_1, H_2, f \}=L(f),$$
where
$$L=L_1 \frac{\partial}{\partial x} + L_2 \frac{\partial}{\partial y} +
L_3 \frac{\partial}{\partial z} \in Vect(X),$$
and
$$L_1=\frac{\partial(H_1,H_2)}{\partial(y,z)},~L_2=\frac{\partial(H_1,H_2)}
{\partial (z,x)},~L_3=\frac{\partial(H_1,H_2)}{\partial(x,y)}.$$
Denote by
$$\tilde{L}=L + \frac{\partial}{\partial t} \in Vect(\tilde{X})$$
corresponding vector field on the extended phase space $\tilde{X}$.

{\bf Theorem 6} {\it Vector field $\tilde{L} \in Vect(\tilde{X})$
is a line field of the $3$-form $ d\omega^{(2)}$, i.e.
$$ i_{\tilde{L}}(d\omega^{(2)})=0.$$}

{\it Proof.} Since
\begin{eqnarray*}
d \omega^{(2)} & = & dx \wedge dy \wedge dz -dH_1 \wedge dH_2 \wedge dt \\
& = & dx \wedge dy \wedge dz - L_3 dx \wedge dy \wedge dt +
L_2 dx \wedge dz \wedge dt -L_1 dy \wedge dz \wedge dt,
\end{eqnarray*}
direct calculation (cf.\ \cite{Ar}) shows that
\begin{eqnarray*}
i_{\tilde{L}}(d\omega^{(2)}) & = & d \omega^{(2)}(\tilde{L},~.~,~.~)=
L_1 dy \wedge dz - L_2 dx \wedge dz + L_3 dx \wedge dy \\
& - & L_3 (L_1 dy \wedge dt - L_2 dx \wedge dt) +
L_2 (L_1 dz \wedge dt - L_3 dx \wedge dt) \\ &  - & L_1 (L_2 dz \wedge dt -
L_3 dy \wedge dt) - L_3 dx \wedge dy + L_2 dx \wedge dz -L_1 dy \wedge dz =0.
\end{eqnarray*}
$\Box$

As in the case of Hamiltonian mechanics, this theorem has important
corollary (cf.\ \cite{Ar}). Namely, let $c$ be a closed $2$-chain in
$\tilde{X}$, $g^t$ be Nambu-Hamilton phase flow and
$3$-chain $J^tc=\{g^{\tau}c,~0 \leq \tau \leq t \}$ be a trace of the chain
$c$ under the isotopy $g^{\tau}$.

{\bf Corollary}
$$\int_{c}\omega^{(2)}=\int_{g^{t}(c)}\omega^{(2)}.$$

{\it Proof.} Using the formula $\partial (J^tc)=c-g^{t}(c)$, Stokes
theorem and Theorem 6, we get
$$\int_{c}\omega^{(2)} - \int_{g^{t}(c)}\omega^{(2)}=\int_{J^tc}d \omega^{(2)}=
0.$$ $\Box$

Next we formulate the principle of least action.
Recall that in Hamiltonian mechanics this principle
states (see, e.g., \cite{Ar}) that classical
trajectory---integral curve $\gamma$ of Hamilton's phase flow with
initial and final points $(p_0,q_0,t_0)$ and
$(p_1,q_1,t_1)$, is an extremal of the action functional
$$A(\gamma)=\int_{\gamma}(pdq - Hdt)$$
in the class of all paths connecting initial and final points in
given $n$-dimensional subspaces $(t=t_0,q=q_0)$ and $(t=t_1,q=q_1)$ in the
extended phase space.

{\bf Definition 5} The functional
$$A(C_{2})=\int_{C_2} \omega^{(2)}$$
---integral of the action form over $2$-chains in the extended phase space
$\tilde{X}$, is called {\it action functional} for Nambu mechanics.

Let $\gamma$ be a closed $1$-chain in $X$. Define
a $2$-chain $\Gamma$ in $\tilde{X}$ as a trace of $\gamma$ under the isotopy
$g^{t},~t_0 \leq t \leq t_1$, so that $\partial \Gamma = \gamma_{t_0}-
\gamma_{t_1}$,
where $\gamma_{t_0}=\gamma,~ \gamma_{t_1}=g^{t_1 - t_0}(\gamma)$.
The following theorem states the principle of least action for Nambu mechanics.

{\bf Theorem 7} {\it A $2$-chain $\Gamma$ is an extremal of the action $A(C)$
in the class of all $2$-chains $C$ whose boundaries---$1$-chains $c_{t_0}$
and $c_{t_1}$---have the same projections onto the $yz$-planes
as a given $1$-chains $\gamma_{t_0}$ and $\gamma_{t_1}$.}

{\it Proof.} As in the case of Hamiltonian mechanics,
this statement follows from the previous theorem (cf.\ \cite{Ar}).

Here is another proof, based on direct calculation. Assume that $\Gamma$
admits a parametrization $x=x(s,t),~
y=y(s,t),~z=z(s,t)$, where $0 \leq s \leq 1,~ t_0 \leq t \leq t_1,$ such that
it time-slices are closed curves, i.e.\ for all $t$ functions $x,y,z$ are
periodic in $s$ with period $1$. Variations $\delta x,~ \delta y, ~
\delta z$ satisfy conditions $ \delta y(s,0)=\delta y(s, 1)=\delta z(s,0)=
\delta z(s,1)=0$ for all $0 \leq s \leq 1$. We have explicitly
$$A(C)=\int \int_{\Pi}\{ x(\frac{\partial y}{\partial s}
\frac{\partial z}{\partial t} -
\frac{\partial y}{\partial t} \frac{\partial z}{\partial s})
 -  H_1 (\frac{\partial H_2}{\partial x} \frac{\partial x}{\partial s} +
\frac{\partial H_2}{\partial y} \frac{\partial y}{\partial s} +
\frac{\partial H_2}{\partial z} \frac{\partial z}{\partial s}) \} dsdt,$$
where $\Pi=\{0 \leq s \leq 1,~t_0 \leq t \leq t_1 \}$.
Using Stokes theorem and properties of $\delta y$ and
$\delta z,$ we obtain the following expression for the variation of the action
\begin{eqnarray*}
\delta A(\Gamma)  =  \int \int_{\Pi} \{
&(&\{\frac{\partial z}{\partial t} - \frac{\partial (H_1, H_2)}{\partial
(y,z)}\}
\frac{\partial y}{\partial s}
 -  \{\frac{\partial y}{\partial t} -
\frac{\partial (H_1, H_2)}{\partial (z,x)}\} \frac{\partial z}{\partial s})
\delta x \\
 +  &(&\{\frac{\partial x}{\partial t} - \frac{\partial (H_1, H_2)}
{\partial (y,z)}\} \frac{\partial z}{\partial s}
 -  \{\frac{\partial z}{\partial t}
- \frac{\partial (H_1, H_2)}{\partial (x,y)}\}
\frac{\partial x}{\partial s}) \delta y \\
+ &(& \{\frac{\partial y}{\partial t} - \frac{\partial (H_1, H_2)}
{\partial (z,x)}\} \frac{\partial x}{\partial s}
 - \{\frac{\partial x}
{\partial t} - \frac{\partial (H_1, H_2)}{\partial (y,z)}\}
\frac{\partial y}{\partial s})\delta z \}dsdt,
\end{eqnarray*}
which shows that $\delta A(\Gamma)=0$ for all admissible variations
$\delta x,~\delta y,~\delta z$, if $\Gamma$ consists of integral curves of
Nambu-Hamilton phase flow. $\Box$

{\bf Remark 8} The converse statement is also true: extremals of
action functional are ``world-sheets'' $x(s,t),y(s,t),z(s,t)$
consisting of families of integrals curves of Nambu-Hamilton
phase flow parametrized by $s$. Indeed, these extrema are characterized by
condition the the cross product of the following vectors in $\R^3$
$$(\frac{\partial x}{\partial t}-
\{H_1,H_2,x\},\frac{\partial y}{\partial t}-\{H_1,H_2,y\},\frac{\partial z}
{\partial t}-\{H_1,H_2,z\})~
{\rm and}~
(\frac{\partial x}{\partial s}, \frac{\partial y}{\partial s},
\frac{\partial z}{\partial s})$$
is zero, which means that there exists a function
$\alpha(s,t)$ such that
\begin{eqnarray*}
\frac{\partial x}{\partial t}-\alpha(s,t) \frac{\partial x}{\partial s}
& = &\frac{\partial(H_1,H_2)}{\partial(y,z)},\\
\frac{\partial y}{\partial t} -\alpha(s,t) \frac{\partial y}{\partial s}
& = &\frac{\partial(H_1,H_2)}{\partial(z,x)},\\
\frac{\partial z}{\partial t}-\alpha(s,t) \frac{\partial z}{\partial s}
& = &\frac{\partial(H_1,H_2)}{\partial(x,y)}.
\end{eqnarray*}
Changing the parametrization $s \mapsto s^{\prime}=s + \alpha(s,t) t,~
t \mapsto t^{\prime}=t$, we see that this system reduces to the family of
Nambu-Hamilton equations parametrized by $s$.

{\bf Remark 9} Comparison between principles of least action for
Nambu and Hamiltonian mechanics shows that ``configuration space'' of
Nambu mechanics of order $3$ constitute ``two-third'' of a phase space.

It is straightforward to generalize presented results for the case of Nambu
bracket of order $n$. The analog of Poincar\'{e}
-Cartan integral invariant is defined as the following $n-1$-form
$$\omega^{(n-1)}=x_1 dx_2 \wedge \ldots \wedge dx_n -H_1 dH_2 \wedge
\ldots \wedge dH_{n-1} \wedge dt,$$
and the action functional is given by
$$A(C_{n-1})=\int_{C_{n-1}} \omega^{(n-1)}$$
and is defined on the $n-1$-chains in the extended phase space.
In its formulation admissible variations are those which do not change
projections of the boundary $\partial C_{n-1}$ on the
$x_2x_3 \ldots x_n$-hyperplanes; in this case the ``share'' of
``configuration space'' in a phase space is $1 - 1/n$.

{\bf Remark 10} This construction of action form and action functional
is somewhat similar to the construction of cyclic cocycles in Connes'
approach to the non-commutative differential geometry \cite{Connes}.

\section{Quantization}

There exist several different (and, in a certain sense, equivalent)
points of view on quantization problem.

One is based on the approach which uses a {\it deformation theory} of
associative algebras. It considers quantization as a deformation
of a (commutative) algebra of classical observables on the phase space in
the ``direction'' defined by a given Poisson (or symplectic) structure
\cite{FBS}.

Namely, let $X$ be a Poisson manifold with the Poisson bracket $\{~,~\}$
and algebra of classical observables $A=C^{\infty}(X)$.
One-parameter family $\{A_h\}$ of associative algebras
is called a {\it quantization} of a commutative algebra $A$ of
classical observables, if following conditions are satisfied.

1. Algebra $A$ is included into this family
$$A_h|_{h=0}=A,$$
(or, in a formal algebraic category, $A \cong A_h/hA_h$).

2. All algebras $A_h$ are isomorphic to $A$ (or rather to $A[[h]]$ in formal
algebraic category) as linear spaces.

3. Denoting by $*_h$  (associative) product in $A_h$ ($*$-product in the
terminology of \cite{FBS}), one has the expansion
$$f_1 *_h f_2  =f_1 f_2 +\frac{h}{2}C_2(f_1,f_2) + O(h^2),$$
with the property
$$C_2(f_1,f_2)-C_2(f_2,f_1)=\{f_1,f_2\}.$$

The last property is often referred as a {\it correspondence principle}
$$\{f_1,f_2\}=\lim_{h \rightarrow 0}\frac{1}{h}(f_1*_h f_2 - f_2*_h f_1)$$
between classical and quantum mechanics.

Deformation theory approach can be carried out explicitly in the case
$X=\R^2$ (or $\R^{2n}$) with canonical Poisson bracket and yields
celebrated {\it Hermann Weyl's quantization} scheme.
Corresponding $*$-product $*_h$---a map from $A \otimes A$ into $A$, is given
by a composition of a usual commutative point-wise product on $A$ with
the following bilinear pseudo-differential operator
$$\exp(\frac{h}{2}\{~,~\})=\exp(\frac{h}{2}\{\frac{\partial}{\partial x} \wedge
\frac{\partial}{\partial y}\}): A \otimes A \mapsto A \otimes A.$$

One might use the same approach towards the quantization of Nambu
mechanics. Namely, suppose that we try to include ``usual''
ternary product---a map from $A \otimes A \otimes A$ into $A$, given by a
point-wise multiplication of classical observables, into the ``new'' ternary
operation $(~,~,~)_h$ which depends on parameter $h$ and satisfies certain
natural properties (analogous to associativity condition in the binary
case) and a correspondence principle
$$\{f_1,f_2,f_3\}=\lim_{h \rightarrow 0}\frac{1}{h}{\rm
Alt}(f_1,f_2,f_3)_{h},$$
where ${\rm Alt}$ denotes complete anti-symmetrization with respect to the
arguments $f_1,f_2,f_3$.

We can easily generalize Weyl's formula by letting the new ternary
operation $(,~,~,~)_h$ to be the composition of a point-wise ternary product
and the following trilinear pseudo-differential operator
$$\exp(\frac{h}{6}\{~,~,~\})=\exp(
\frac{h}{6}\{\frac{\partial}{\partial x} \wedge \frac{\partial}
{\partial y} \wedge \frac{\partial}{\partial z}\}):
A \otimes A \otimes A  \mapsto A \otimes A \otimes A.$$
This formula generalizes verbatim for canonical Nambu bracket of order $n$.

{\bf Remark 11} One should note the following.

(i) Contrary to the usual binary case, point-wise ternary product of
classical observables looks less natural.

(ii) ``Natural'' constraints for ternary operations generalizing
usual associativity constraint are, apparently, not well understood.

Nevertheless, proposed deformation has certain ``appeal'' and should be
analyzed further. It might lead to the (partial) answer to a problem in (ii).

{\bf Remark 12} Recently R.~Lawrence \cite{Lawrence} proposed a system of
axioms for linear spaces with $n$-ary operations ($n$-algebras using her
terminology). It looks that ``deformed'' product $(~,~,~)_h$, as well
as a ternary point-wise product, do not satisfy them. However,
one should have in mind that approach in
\cite{Lawrence} generalizes certain combinatorial properties of
algebraic operations (notably the Stasheff polyhedron), whereas our approach
is based on ``dynamics''.

Another approach uses {\it Feynmann's path integral formulation} of
quantum mechanics. For a quantum particle, described classically by a
Hamiltonian system, it gives the probability amplitude of the
transition from a state $|q_0\rangle$ at time
$t=t_0$ into state $|q_1\rangle$ at time $t=t_1$ as a functional integral
of the exponential of the classical action
$$\exp\{ \frac{i}{h}A(\gamma)\}$$
over all ``histories'' $\gamma$ with respect to the ``Liouville measure''
$$\frac{i}{h}DpDq$$
in the functional space of all ``histories''.

Principle of the least action for Nambu mechanics can be used
to formulate just similar rule.  However, special form of action principle
(see Theorem 7) requires that quantum states $|y(s),z(s)\rangle$ should be
parametrized by loops $y(s),z(s)$ rather then by points in a ``configuration''
space. This natural appearance of loops looks quite appealing. It suggests that
one might still have particles as a point-like objects in classical
picture and dynamic of loops (or, generally, $n-2$-closed chains) in
quantum picture. We are planning to investigate this intriguing opportunity
elsewhere.

Finally, we present yet another approach to quantization
a {\it canonical formalism}. It is based on {\it Heisenberg commutation
relations}, which for the phase space $X=\R^2$ with canonical Poisson bracket
look the following (in a complex form)
$$[a, a^{\dagger}]= a a^{\dagger} - a^{\dagger} a =I,$$
where operators $a^{\dagger},~a$ act in a linear space of quantum states.
They have the following
realization in a space ${\bf H}_2$ with the basis $\{|n\rangle\}_{n \geq 0}$
parametrized by non-negative rational integers:
$$a|n\rangle=\sqrt{n}|n-1\rangle,~a^{\dagger}|n\rangle=\sqrt{n+1}|n+1\rangle.$$
The vector $|0\rangle$ plays the role of a vacuum state and operators
$a^{\dagger},~a$ are called creation-annihilation operators.

Being one of the fundamental principles of quantum mechanics,
Heisenberg commutation relations have remarkable mathematical properties.
In particular, one has celebrated Stone-von Newmann theorem that all
irreducible representations of Heisenberg commutation relations are unitary
equivalent. Application of Stone-von Newmann theorem  to the linear canonical
transformations gives a certain projective representation of the symplectic
group, which was shown by Andr\'{e} Weil \cite{Weil} to play a fundamental
role in the arithmetics of quadratic forms and quadratic reciprocity for
number fields.

In \cite{Nambu} Nambu proposed the following generalization of Heisenberg
commutation relation
\beq \label{NH}
[A_1,A_2,A_3]=A_1A_2A_3 - A_1A_3A_2 + A_3A_1A_2 - A_3A_2A_1 + A_2A_3A_1 -
A_2A_1A_3 =cI,
\eeq
where $A_1,A_2,A_3$ are linear operators, $I$ is
a unit operator and $c$ is a constant. We will call (\ref{NH})
{\it Nambu-Heisenberg relation}.

{\bf Remark 13} Although Nambu-Heisenberg relation
satisfies correspondence principle
$$\frac{1}{h}[~,~,~] \mapsto \{~,~,~\},$$
{\it a priori} it is not clear that observables in quantum Nambu mechanics
should be realized by linear operators. Therefore the form (\ref{NH}) might
look rather {\it ad hoc}. It was already pointed by Nambu \cite{Nambu} that,
contrary to the usual commutator $[~,~]$,
``ternary commutator'' $[~,~,~]$ is not a derivation with respect to the
operator product, which poses certain problems in formulating
quantum dynamics (see \cite{Nambu} for detailed discussion).
{}From our point of view, it looks like
there are no reasons to impose such derivation property
with respect to the operator product. ``Triple commutator''
$[~,~,~]$, should be considered to define a Nambu-Lie ``gebra'' structure on
quantum observables. Its property of being a derivation of this
structure (see Definition 3) will imply consistency of quantum Nambu dynamics.

Instead of continued this discussion (which we are planning to do
elsewhere), here we will only present
explicit realization of Nambu-Heisenberg relation (\ref{NH}).

Let ${\bf Q}[\rho]$ be a quadratic number field with
$1+\rho + \rho^2=0$, i.e.
$$\rho=\frac{-1 + \sqrt{-3}}{2},$$
and let ${\bf Z}[\rho]$ be a ring of algebraic integers in ${\bf Q}[\rho]$,
i.e.
$$\omega = m_1 + m_2 \rho \in {\bf Z}[\rho],~m_1,m_2 \in {\bf Z}.$$
Denote by ${\bf H}_3$ linear space with the basis $\{|\omega\rangle\}$
parametrized by ${\bf Z}[\rho]$.
Direct calculation proves the following result.

{\bf Theorem 8} {\it Nambu-Heisenberg relation with $
c=\rho - \bar{\rho}=\sqrt{-3}$ admits the
following representation
$$A_1|\omega\rangle=(\omega + 1 + \rho)|\omega +1\rangle,~ A_2|\omega\rangle=
(\omega + \rho)|\omega + \rho\rangle,~A_3|\omega\rangle=
\omega |\omega + \rho^2\rangle$$
in the space ${\bf H}_3$.}

The following arguments explain why such type
of reprentations are possible.

Using  the analogy with $n=2$ case, consider representation (\ref{NH})
of the form
$$A_1|\omega\rangle=f_1(\omega)|\omega +1\rangle,~A_2|\omega\rangle=
f_2(\omega)|\omega + \rho\rangle,~ A_3|\omega\rangle
=f_3(\omega)|\omega + \rho^2\rangle.$$
Because of $1 + \rho + \rho^2=0$, vectors $|\omega\rangle$ are eigen-vectors
for all possible triple products $A_{i_1}A_{i_2}A_{i_3}$ of $A_1,A_2,A_3$
and Nambu-Heisenberg relation (\ref{NH})
reduces to functional equation
for the uknowns $f_1(\omega),~f_2(\omega),~f_3(\omega)$. Theorem 8 exhibits
one of its solutions. We can obtain other solutions using the following trick.
Equation (\ref{NH}) is invariant under the similarity transformation
$$A_i \mapsto \tilde{A}_i=U^{-1}A_iU,~i=1,2,3,$$
where $U: {\bf H}_3 \mapsto {\bf H}_3$ is an invertible linear operator.
Choose $U$ to
be diagonal in the standard basis of ${\bf H}_3$, i.e.
$$U|\omega\rangle=u(\omega)|\omega\rangle,~ u(\omega) \neq 0,~
\omega \in {\bf Z}[\rho].$$
Then operators $\tilde{A}_i$ are represented by the same formulas as $A_i$'s
with
$$\tilde{f}_1(\omega)=u^{-1}(\omega + 1)u(\omega)f_1(\omega),~ \tilde{f}_2
(\omega)=u^{-1}(\omega + \rho)u(\omega)f_2(\omega),~\tilde{f}_3(\omega)=
u^{-1}(\omega + \rho)u(\omega)f_3(\omega).$$
Now assume that $A_1$ and $A_2$ commute and
$f_1,f_2$ are non-zero for all $\omega \in {\bf Z}[\rho]$. Then
one can choose $u(\omega)$ such that $\tilde{f}_1(\omega)=
\tilde{f}_2(\omega)=1$. Indeed, $u(\omega)$
can be found as a solution of the following compatible system (because of the
commutativity of $A_1$ and $A_2$) of difference
equations
\begin{eqnarray*}
u(\omega +1) & = & f_1(\omega)u(\omega), \\
u(\omega + \rho) & = &f_2(\omega)u(\omega).
\end{eqnarray*}
(Provided certain conditions at ``infinity'' for the
functions $f_1$ and $f_2$ are satisfied, $u(\omega)$ can be given as
semi-infinite product over the lattice ${\bf Z}[\rho]$.). Therefore,
original functional equation for $f_i$'s reduces
to the following simple equation
$$\tilde{f}_3(\omega+1)-\tilde{f}_3(\omega)=c,$$
which can be easily solved.

``Gauging back'' these solutions with $\tilde{f}_1=\tilde{f}_2=1$
we obtain other solutions.

Solution presented in Theorem 8 does not belong to this class:
operators $A_1$ and $A_2$ (as well as other pairs) do not commute and, since
vectors
$|\rho^2\rangle,~|1+\rho^2\rangle$ and $|0\rangle$ annihilate,
operators $A_1,A_2$ and $A_3$, corresponding functions $f_i$ have zeros.
This implies that in this case corresponding invertible operator $U$ does not
exist. This shows the special role played by representation in Theorem 8:
although there is no vacuum vector in a conventional sence,
there exists triple of vectors
$$|\rho^2\rangle,~|1+\rho^2\rangle,~|0\rangle$$
with ``similar'' properties.

{\bf Remark 14} It should be noted that this realization
of Nambu-Heisenberg relation must be supplemented with the analog of
Hermitian anti-involution, which is this case should be an operation of
order 3 (since we a dealing with a triple instead of a pair). We are
planning to address this issue elsewhere.

Finally, consider briefly the general case of Nambu mechanics of order $n$.
One can postulate the following form of Nambu-Heisenberg relation
\beq \label{GNH}
\sum_{\sigma \in Symm(n)}(-1)^{\epsilon(\sigma)}A_{\sigma(1)} \ldots
A_{\sigma(n)}=cI.
\eeq
Assuming that $n$ is a prime integer greater then $2$,
consider representation of (\ref{GNH}) in a linear space ${\bf H}_n$ with
the basis $\{|\omega\rangle\}$ parametrized by algebraic integers
${\bf Z}[\rho]$ in the cyclotomic field of $n$-th root of unity, i.e.
$\rho^n=1$, or $$1 + \rho + \ldots + \rho^{n-1}=0.$$
Assuming that operators $A_i$ are represented in a similar way:
$$A_i|\omega\rangle=f_i(\omega)|\omega + \rho^{i-1}\rangle,~i=1, \ldots, n,$$
one might try to repeat the same trick: ``gauge out'' the
coefficients $f_1, \ldots, f_{n-1}$. Namely, assume that
$f_i$'s are such that the following system of equations
$$u(\omega + \rho^{i-1})=f_i(\omega)u(\omega),~i=1, \ldots n-1$$
is compatible and
has a solution which gives rise to an invertible operator $U$.
In this case commuting
operators $\tilde{A}_i=U^{-1}A_iU$ for $i=1, \ldots, n-1,$ will
represent $n-1$ basic translations (generators) of the lattice ${\bf Z}[\rho]$
and, therefore.
However, analyzing relation (\ref{GNH}) for general $n$
we see that all its terms contain at least one pair of $A_i$'s with
$i \leq n-1$ as nearest
neighbors. These terms will appear twice in (\ref{GNH}) and
commutativity of $\tilde{A}_i$'s implies that they will mutually cancel each
other!

This shows that in the higher order case Nambu-Heisenberg relation
does not admit simple solutions with $n-1$ commuting
operators $A_i$'s. However, we suspect (and are planning to investigate it)
that it has solution similar to that in Theorem 8.

\section{Conclusion}

Though this paper poses more questions that provides the answers, we feel that
the subject of higher order algebraic operations might be relevant for
future development of mathematical structures related to physical problems.
We suspect that it might clarify certain problems related to the
generalizations of the integrability concept (Yang-Baxter equation, Poisson-Lie
groups, quantum groups) for the higher dimensional case (Zamolodchikov's
tetrahedron equations, Frenkel-Moore solutions, $2$-categories).
Ending on highly speculative note, one might suggest that perhaps these higher
order structures can be also relevant in the arithmetics of forms of higher
degrees, higher reciprocity laws in number theory, etc.

\end{document}